\documentclass[aps,prl,twocolumn,footinbib,reprint]{revtex4}
\usepackage[T1]{fontenc}
\usepackage[latin9]{inputenc}
\usepackage{amsmath}
\usepackage{amssymb}
\usepackage{graphicx}
\usepackage{esint}

\usepackage{color}
\usepackage[normalem]{ulem}



\begin{document}

\title{Scale and Nature of  Sulcification Patterns}

\author{Evan Hohlfeld}
\email{evanhohlfeld@gmail.com}
\affiliation{Lawrence Berkeley National Lab, Berkeley, California 94720, USA}
\author{L. Mahadevan}
\email{lm@seas.harvard.edu}
 \affiliation{Department of Physics, and School of Engineering and Applied Sciences,   Harvard University, Cambridge, Massachusetts 02138, USA}

\begin{abstract}
Sulci are surface folds commonly seen in strained soft elastomers and form via a strongly subsubcriticalcritical, yet scale-free instability. Treating the threshold for nonlinear instability as a  nonlinear critical point, we explain the nature of sulcus patterns  in terms of the scale and translation symmetries which are broken by the formation of an isolated, small sulcus. Our perturbative theory and simulations show that  sulcus formation in a thick, compressed slab can arise either as a supercritical  or as a weakly subcritical bifurcation relative to this nonlinear critical point, depending on the boundary conditions. An infinite number of competing, equilibrium patterns simultaneously emerge at this critical point, but the one selected has the lowest energy. We give a simple, physical explanation for the formation of these sulcification patterns using an analogy to a solid-solid phase transition with a finite energy of transformation.
\end{abstract}

\maketitle

Elastic pattern forming instabilities such as wrinkling typically have an intrinsic wavelength or scale
set by a combination of geometric and material parameters.  Patterns of sulci, which are sharply creased self-contacting folds \cite{Biot, TTea88, ANG99,AGALD,VTJK,P.-M.-Reis:2009fk, Hayward}, are fundamentally different in that they result from a nonlinear instability with no intrinsic scale \cite{Hohlfeld-Mahadevan}. Therefore, the question of scale and pattern selection for sulcus patterns naturally arrises.
Previously, we explained that while sulci may nucleate and grow when a short-wavelength \emph{linear} instability sets in at a point on a free surface where a critical compression is achieved, the instability is actually  subcritical \cite{Hohlfeld-Mahadevan} as there is a second lower critical compression at which the \emph{linearly stable} surface can develop a sulcus of vanishing size. At larger values of compression the surface is \emph{metastable}---an infinitesimal sulcus can nucleate and grow, but the surface remains linearly stable. 
The lower and higher critical compressions are similar to the binodal and spinodal points of the liquid-vapor transition in fluids. 
Growing interest in the ramifications of this instability which can be driven by by swelling \cite{TTea88,VTJK}, by mechanical deformation \cite{ANG99,AGALD,P.-M.-Reis:2009fk,Hohlfeld-Mahadevan},  growth or  morphogenesis \cite{Benamar, Suo}, or an applied  field \cite{Zhao} has focused on characterizing the sensitivity of sulcification to defects \cite{Hutchinson},  and  on control \cite{Hayward,Zhao}. However, there is as yet no understanding of the behavior of a sulcus or patterns thereof near the onset of the instability at the lower critical strain, nor a general theory of such scale-free instabilities. 

In this Letter, we address both these issues by considering the asymptotic structure of both the deformation and the energy of sulcus patterns near the threshold of the instability in homogeneously strained slabs. We find that sulcification is always deeply subcritical relative to  Biot's threshold compression and has no weakly supercritical regime. However, the instability and concomitant bifurcation can be either supercritical or weakly subcritical relative to the recently discovered lower critical compressive strain  \cite{Hohlfeld-Mahadevan}, depending on the boundary conditions.
As such, the instability at the smaller critical compression can be viewed as a new kind of critical point---one unrelated to a linear instability---and we introduce a perturbation theory based on the arbitrary scale and position
 of an infinitesimal sulcus that forms at this critical point. Henceforth,  the terms sub- and supercritical will 
 refer to the type of instability relative to 
 this critical point.
 Our simulations and theory predict 
 a variety of equilibrium patterns, related to the nearly arbitrary number, locations, and sizes of incipient sulci. We also calculate the  energy  and the depths of the folds within a pattern by linearizing about the critical configuration prior to sulcification.  

 \begin{figure}
\includegraphics[width=1\columnwidth]{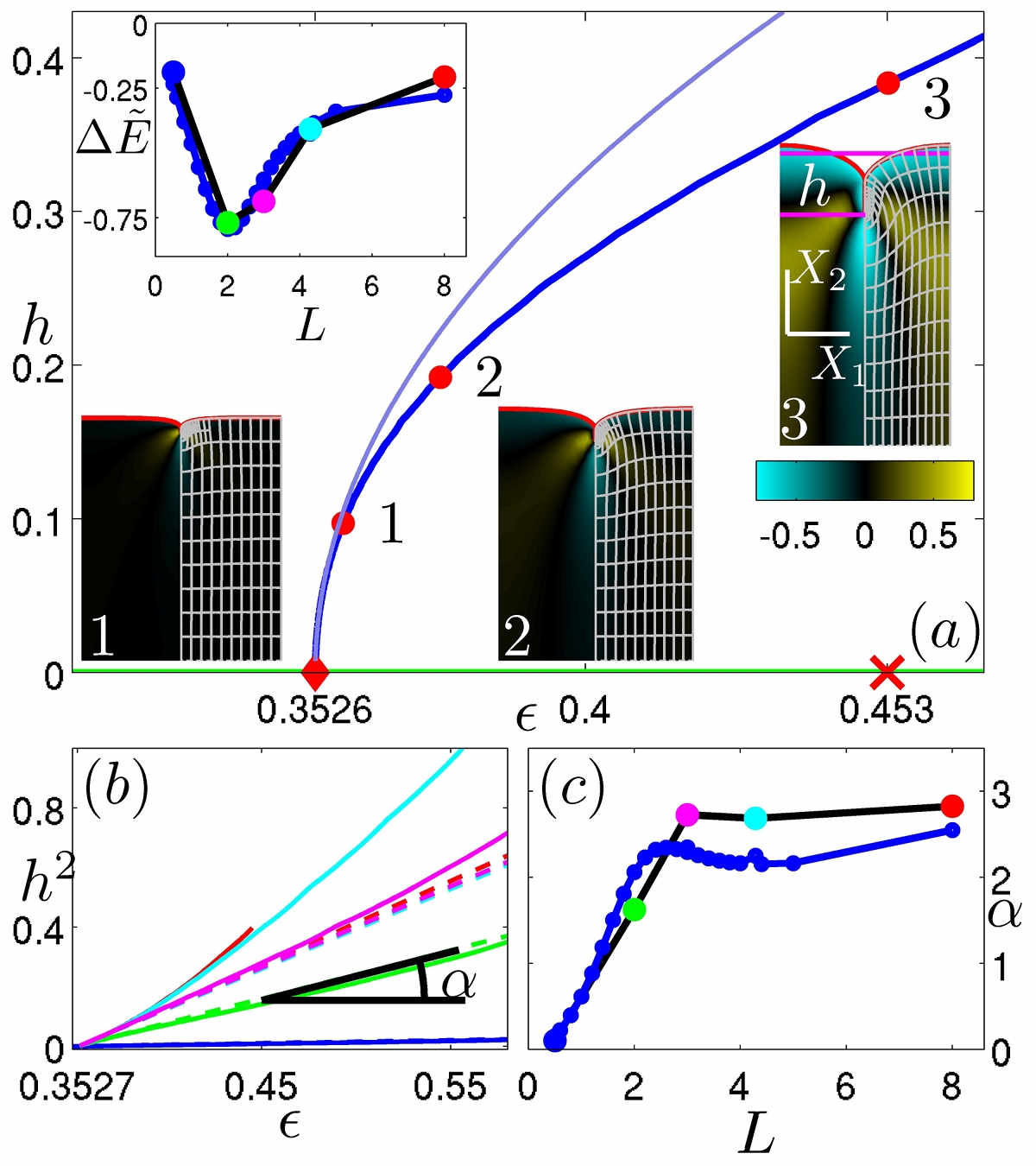}
\caption{\label{fig:fig1}{(a) Periodic sulcified deformations of a freely sliding,
supported slab with pattern wavelength $L=2$; unit cells labeled $1-3$  correspond to labeled points in the bifurcation diagram (simulation: dark blue, leading order theory: light blue, reference configuration: green).
In the deformations, $h$ is the relative depth of a sulcus, the deformed Lagrangian mesh is overlaid in light gray and color indicates $W(\partial\mathbf{x}_{s,L}/\partial \mathbf{X})-W(\partial\mathbf{x}_r/\partial \mathbf{X})$. The top magenta line is the surface of $\mathbf{x}_r$; the red line is the surface of $\mathbf{x}_{s,L}$. 
Biot's theshold is
 $\epsilon\approx43.5\%$ {(red $\times$).} Inset $\Delta \tilde{E}$ [see Eq. (\ref{eq:reduced_energy})] [simulation: black, theory: blue, marker colors correspond with (b)] (b)  Bifurcation diagrams for supported slabs (Red, $L=8$; cyan, $L=4.29$; magenta, $L=3$; green, $L=2$; and blue, $L=0.5$); dashed lines are linear fits to the near-threshold region. (c) $\alpha$  [see Eq. (\ref{eq:h-eps-relation})] (simulation: black curves, marker colors correspond with (b), theory: blue line).}
}
\end{figure}

We consider a thick, two-dimensional, incompressible soft elastomer slab subject to uniform compression along the $X_1$-axis [see Fig. (\ref{fig:fig1}a)], with nominal stretch $1-\epsilon$. We consider both supported slabs which are free to slide along the $X_1$-axis and unsupported slabs. In both cases we assume reflection symmetry about the sides parallel to the $X_2$-axis.
 A simple energy functional for planar deformations $\mathbf{x}\left(\mathbf{X}\right)$ of any such slab  (with reference body $\Omega\subset\mathbb{R}^{2}$) is 
\begin{equation}
E\left(\mathbf{x}\right)= \int_{\Omega }W\left(\frac{\partial\mathbf{x}}{\partial \mathbf{X}}\right)d^{2}\mathbf{X}  = \int_{\Omega}\frac{1}{2}\left(\left| \frac{\partial\mathbf{x}}{\partial \mathbf{X}}\right|^{2}-2\right)d^{2}\mathbf{X}\label{eq:neo-hookean-energy}
\end{equation}
subject to the \emph{nonconvex} constraint $\det(\partial \mathbf{x}/\partial \mathbf{X})=1$ and the topological condition that $\mathbf{x}\left(\mathbf{X}\right)$ is globally
invertible \footnote{
Reflection symmetry simplifies the invertibility constraint 
to a 
unilateral contact problem within the fold of a sulcus}. Energy (\ref{eq:neo-hookean-energy}) corresponds to the classical neo-Hookean model 
with a unit shear modulus,
and $\partial\mathbf{x}/\partial\mathbf{X}$ is the deformation gradient $\mathbf{A}$.  We denote the critical deformation gradient  by $\mathbf{A}_{c},$ which corresponds to a surface-parallel compression of magnitude $\epsilon_c=35.3\%.$

For both supported and unsupported slabs, we conducted finite element simulations similar to those described in \cite{Hohlfeld-Mahadevan}. Our simulations used a  mesh that was recursively refined in the neighborhood of an incipient crease singularity (where the surface normal changes sign at the bottom  of a sulcus) so that the local mesh scale relative to the thickness of the slab (assumed to be unity)  is  $l_{m}=O\left(10^{-5}\right)$. To eliminate the effects of the mesh on small sulci we added 
a
numerical regularization to 
energy (\ref{eq:neo-hookean-energy}) in the form of a surface bending energy (with zero surface tension), setting a regularization scale $l_{r}=O\left(10^{-4}\right)$.
Our simulation results were unchanged when we increased $l_{m}$ and $l_{r}$ by an order of magnitude.
Slabs of the neo-Hookean material compressed to $\epsilon>\epsilon_c$ are energetically unstable---with an energy barrier for sulcification that vanishes as $l_r\rightarrow 0$ \cite{Hohlfeld-Mahadevan}. To calculate the pattern of sulcification in strained slabs we used inhomogeneous normal stresses on the surface to create patterned sulcified slabs, removed the normal stress, and then followed the branches of sulcified deformations using  a variant of pseudo arc-length continuation in $\epsilon$ \cite{Hohlfeld-Mahadevan}. We denote uniformly compressed reference configurations as $\mathbf{x}_{r}(\mathbf{X},\epsilon)$ and sulcified deformations with scaled wavelength $L$ (relative to the slab depth) as $\mathbf{x}_{s,L}(\mathbf{X},\epsilon);$ sometimes the dependence on $\mathbf{X}$ and/or $L$ is implicit. We measure the size of a sulcus (i.e. the amplitude of a pattern of identical sulci) by the scaled depth, $h,$ of its crease singularity relative to $\mathbf{x}_{r}(\epsilon)$. 
For a crease at $\mathbf{X}=\mathbf{0},$  $h(\epsilon)\equiv x_{r}(\mathbf{0},\epsilon)-x_{s,L}(\mathbf{0},\epsilon)$.

For supported slabs, our simulation results  with $L=2$ are presented in Fig. \ref{fig:fig1}(a), which shows a bifurcation diagram relating $h$ to $\epsilon$ with accompanying representative equilibria. We find that  there is a supercritical instability towards sulcification at $\epsilon=\epsilon_{c}$, but that there are  no linear instabilities until $\epsilon>\epsilon_{B}=45.3\%$ (red $\times$), 
Biot's
threshold strain for surface instability.   Close to the bifurcation point (red diamond), larger (image 2) and smaller (image 1) sulci in the fixed pattern are geometrically similar, much like an isolated sulcus in a bent slab \cite{Hohlfeld-Mahadevan}. This simple scaling 
of a sulcus close to threshold maintains the size of nonlinearities in $\partial\mathbf{x}/\partial \mathbf{X}$ as the pattern amplitude $h$ vanishes. This means that we cannot apply an ordinary normal form theory to explain sulcus patterns. Bifurcation diagrams  for sulcification patterns with other values of $L$ are shown in Fig. \ref{fig:fig1}(b), and
satisfy
\begin{equation}
h^{2}=\alpha\left(L\right)\left(\epsilon-\epsilon_{c}\right)+o(|\epsilon-\epsilon_c|)\label{eq:h-eps-relation}
\end{equation}
where the coefficient $\alpha$ increases with $L$ to a maximum at  $L\approx2.3$ and then saturates [see Figs. \ref{fig:fig1}(b) and \ref{fig:fig1}(c)]. We find that deformations
with  $L>2.3$ are metastable toward further sulcification on a sublattice shifted by $L/2$; those with $L<2.3$ are not metastable.  By minimizing the energy difference $\Delta E=E\left[\mathbf{x}_{s,L}({\epsilon})\right]-E\left[\mathbf{x}_{r}(\epsilon)\right]$  over $L$ at fixed $\epsilon$ we find 
that sulcification, at threshold, sets in with $L\approx 2$. A 
recent 
numerical and experimental study of sulcification patterns in uniaxially strained supported films \cite{Cai} 
agrees with our analytical prediction of $L\approx2$ (given below)
for near-threshold patterns.
 As $\epsilon$ increases beyond threshold, the sulcus pattern undergoes a sequence of transformations and the Lagrangian wavelength $L$ increases, but  the Eulerian wavelength  holds steady at $(1-\epsilon) L \approx 2.$ Near threshold and for fixed $L$ we find $\Delta E(\epsilon,L)/L\propto(\epsilon-\epsilon_c)^2$, so that the expected pattern at onset should minimize the reduced energy 
\begin{equation}
\Delta \tilde{E}(L) = \lim_{\epsilon\to\epsilon_c}E(\epsilon,L)/L(\epsilon-\epsilon_c)^2,\label{eq:reduced_energy}
\end{equation}
In Fig. (\ref{fig:fig1}a)(inset) we see that  minimizing the reduced energy selects a single pattern; the bifurcation diagram corresponds to the minimum reduced energy (green dot).

\begin{figure}
\includegraphics[width=1\columnwidth]{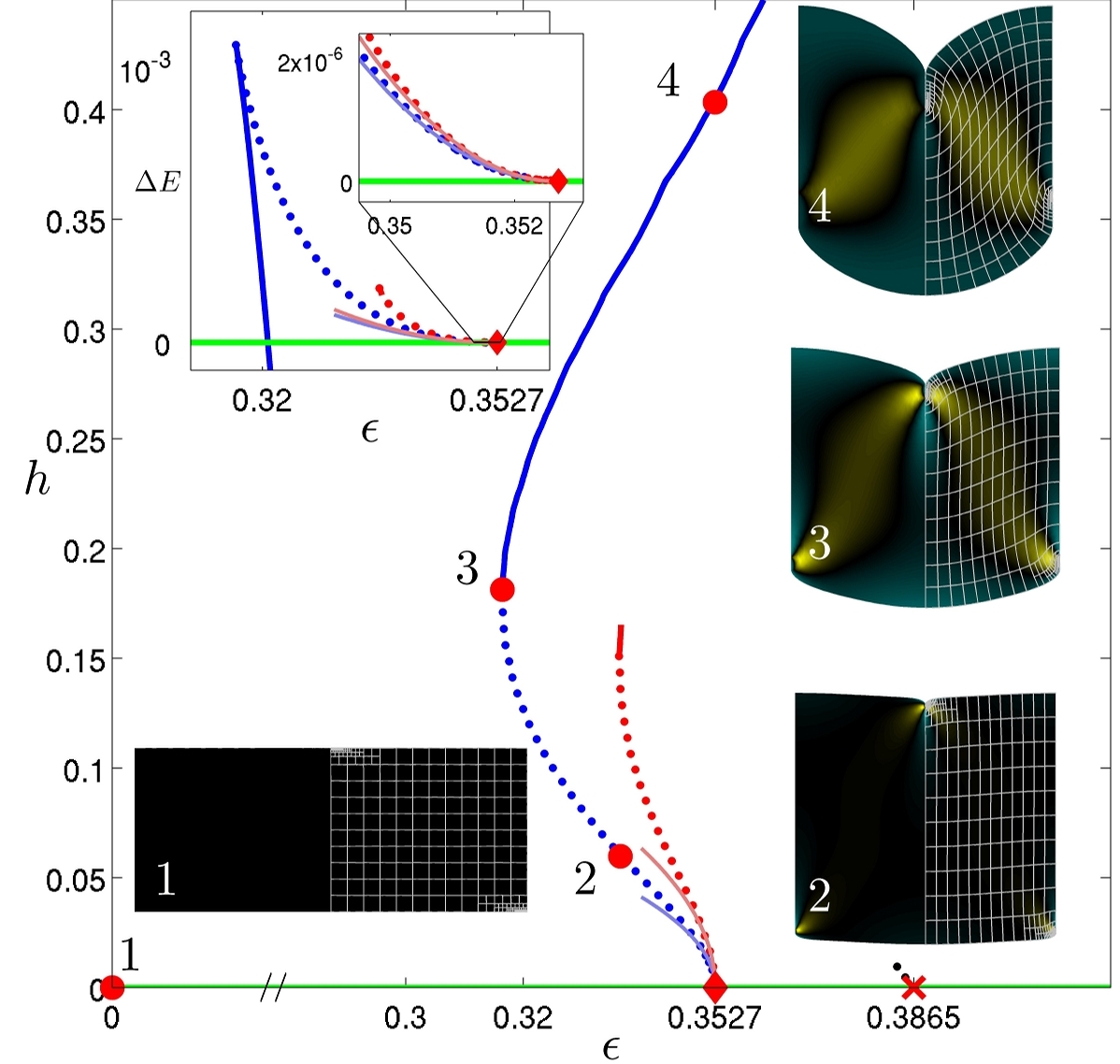}
\caption{\label{fig:fig2}Sulcified deformations of an unsupported slab with a scaled length $L=2.4$; equilibria labeled $1-4$ correspond to labeled points in the bifurcation diagram (simulation: dark blue,  theory: light blue).
 The dotted lines are 
 linearly 
 unstable deformations. The branch of sulcified deformations with one sulcus (simulation: dark red, theory: light red), which also bifurcates at  $\epsilon=35.3\%$ (red diamond), and the branch of smoothly buckled deformations (black), which bifurcates at $\epsilon=38.7\%$ (red $\times$), both terminate when the Biot threshold is reached. Color indicates the change in energy density as in Fig. (\ref{fig:fig1}a), but has been rescaled. Inset: 
 $\Delta E$ as a function of $\epsilon$; colors and markers correspond with the main panel.}
\end{figure} 

For an unsupported slab with scaled length $L=2.4$  we see that sulcification sets in as a subcritical instability;
 see 
 Fig. \ref{fig:fig2}. Unsupported slabs can also buckle, but the critical compression for buckling  approaches $\epsilon_{B}$ as the slab thickness increases (or $L\to 0$) \cite{Biot}. We find two equilibrium patterns, one with sulci on one face of the slab (red line) and another with sulci on both faces of the slab (blue line). 
These bifurcations are subcritical because the formation of a sulcus drives a global flexure mode of the slab (which does not exist in the supported case) and the resulting bending of the slab acts to \emph{increase} the compressive strain at the sulcus. 
The energies of the patterned configurations relative to the energy of $\mathbf{x}_r(\epsilon)$ are shown in
Fig. \ref{fig:fig2} (inset). For linearly unstable configurations, the positive energy difference is the barrier to nucleating the corresponding sulcus pattern; 
the nucleation barrier is lowest for the pattern with sulci on both faces of the slab. Our simulations further show that the pattern with sulci on just one face of the slab is always metastable. The corresponding branch of deformations abruptly terminates as explained in Ref. \cite{Hohlfeld-Mahadevan}, when Biot's threshold strain is attained on the unsulcified surface of the slab. In contrast, the branch of patterns with sulci on both faces continues to all $\epsilon>\epsilon_c$. 
We remark that the uniformly compressed slab (green line) first encounters a linear
buckling
 instability at $\epsilon=38.7\%$ (red $\times$)
  \cite{Biot}. This buckling is subcritical and metastable, and the corresponding branch of deformations (dotted black line) terminates---before reaching a stable configuration---when Biot's threshold strain is achieved on both faces of the slab.

We can explain our simulations and the nature of  sulcification patterns as captured by 
 Eqs. (\ref{eq:h-eps-relation}-\ref{eq:reduced_energy}) 
 using a multiscale  series expansion of near-threshold sulcified solutions to the Euler-Lagrange equation associated with the energy (\ref{eq:neo-hookean-energy}),
 \begin{equation}
\frac{\partial}{\partial X_j}\left(\frac{\partial x_i}{\partial X_j}-p\,\mathrm{cof}_{ij}\left(\frac{\partial \mathbf{x}}{\partial \mathbf{X}}\right)\right)=0,\label{eq:EL}
 \end{equation}
where $p$ is the pressure and $\det(\partial \mathbf{x}/\partial \mathbf{X})=1$. 
  The relevant  scales in this series expansion are the distance to the threshold strain $\delta\equiv\epsilon-\epsilon_c$ and the scaled depths $h_i$ of $N$ sulci with creases at the prescribed points $\{\mathbf{X}_i\}_{i=1}^N$. We will show that for self-consistency $h_i^2\propto\delta$.  In an inner region of size $h_i$ near the $i$ th sulcus, we define inner coordinates $\mathbf{Y}_{i}=(\mathbf{X}-\mathbf{X}_{i})/h_{i}$ and assume {$h_i$ is the only relevant scale within this region, with $\mathbf{x}_{s}(\mathbf{X},\epsilon)-\mathbf{x}_{r}(\mathbf{X},\epsilon_c)=\sum_{n}h_i^n \mathbf{u}_i^{(n)}(\mathbf{Y}_i)$. The deformation in the global outer region is smooth,  so we assume 
  it
  is an analytic function of the various scales, and write  $\mathbf{x}_{s}(\mathbf{X},\epsilon)-x_{r}(\mathbf{X},\epsilon_c)$ as a power series in $\delta$ and $h_1,\dots,h_N$, retaining the terms of the form  $h_i^{n}\delta^m \tilde{\mathbf{u}}_i^{(n,m)}(\mathbf{X})$.  
  (A
   similar series 
   holds 
   for the pressure $p$.)  The multiscale expansion is consistent if the inner and outer solutions match, that is, if they have the same functional form order-by-order in $h_i$ in the intermediate matching regions $|\mathbf{Y}_i|\gg1,\,|\mathbf{X}-\mathbf{X}_i|\lesssim\rho(h_i)$ where the function $\rho(h_i)\to 0,\,\rho(h_i)/h_i\to\infty$ as $h_i\to0^+$.

In the inner region of the $i^{th}$ sulcus $|\partial\mathbf{u}_i^{(0)}/\partial \mathbf{X}|\sim h_i^{-1}$  and so $\mathbf{u}_i^{(0)}=0$, leaving the leading term $\mathbf{v}_s\equiv\mathbf{u}_i^{(1)}$.  Formally, $h_{i}\mathbf{A}_c\cdot\mathbf{Y}_i+h_{i}\mathbf{v}_s(\mathbf{Y}_i)$ is a solution to  the fully nonlinear Euler-Lagrange equation (\ref{eq:EL}) on a traction-free half-space with no self-penetration boundary conditions. It describes an isolated, self-equilibrated sulcus. Because the problem for $\mathbf{v}_s$ is invariant under rescaling, $\mathbf{v}_s(\mathbf{Y}_i)\to \lambda \mathbf{v}_s(\mathbf{Y}_i/\lambda)$ for $\lambda>0$, and under translations, $\mathbf{v}_s(\mathbf{Y}_i)\to\mathbf{v}_s(\mathbf{Y}_i+\mathbf{t})$ where $t_2=0$,
this isolated sulcus can be rescaled or translated to yield an infinity of solutions to the leading order inner problem, a fact that will be significant when we  match the inner and outer solutions.

Computing $\mathbf{v}_s$ numerically showed  that $|\partial\mathbf{v}_s/\partial\mathbf{Y}|$ is bounded, but  the pressure $p(\mathbf{Y}_i)\sim\log(|\mathbf{Y}_i|)$ for $|\mathbf{Y}_i|\ll1,$ in agreement with exact solutions for a crease singularity \cite{Silling}. In the intermediate matching region $\mathbf{v}_s$ has a multipole expansion where the monopole term 
vanishes
because the sulcus is self-equilibrated.  We find that $h_{i}\mathbf{v}_s(\mathbf{Y}_{i})\sim a\mathbf{f}_1(\theta)h_{i}^{2}r_{i}^{-1}+b\mathbf{f}_2(\theta)h_{i}^{3}r_{i}^{-2}+O(h_{i}^4r_{i}^{-3})$ for $|\mathbf{Y}|_i\gg1$ where $r_{i}=|\mathbf{Y}_{i}|$, $\theta$ is the polar coordinate,  the dipole strength $a=-1.17$ and the quadrupole strength $b=1.21$ in a normalization where $\mathbf{f}_1(-\frac{\pi}{2})=\mathbf{f}_2(-\frac{\pi}{2})=\hat{\mathbf{n}}$, the outward surface normal vector. 

In the outer region, because $|\partial \tilde{\mathbf{u}}_i^{(n,m)}/\partial \mathbf{X}|$ is small here, $\tilde{\mathbf{u}}_i^{(n,m)}$ can be computed using perturbation theory and at leading order yields $\tilde{\mathbf{u}}^{(0,1)}=\partial\mathbf{x}_r(\mathbf{X},\epsilon_c)/\partial\epsilon$. Then by matching the dipole  moments  of the inner regions, we find that $\tilde{\mathbf{u}}_i^{(1,0)}=\mathbf{0}$ and  $\tilde{\mathbf{u}}_i^{(2,0)}=a \mathbf{f}_1(\theta)r_i^{-1}+\mathbf{w}_{2i}$, where $\mathbf{w}_{2i}$ is smooth and depends on the domain $\Omega$ and the boundary conditions; $\tilde{\mathbf{u}}_i^{(2,0)}$ is the linear response of the outer region to a force dipole  at $\mathbf{X}_i$.

The leading order match does not involve $\delta$, and so does not fix a unique relationship between $h_i$ and $\delta$. Therefore we look to the next order in $h_i$ to explain the empirical result Eq. (\ref{eq:h-eps-relation}). We find that because of the scale and translational degeneracy of an isolated sulcus,  solutions to the inner and outer problems will only exist if the sulci \emph{move} to sit at local maxima of strain corresponding to the critical value $\epsilon_c$ as the sulci grow. We can state these conditions formally in terms of the residual deformation
\begin{equation}
\mathcal{U}_i(\mathbf{X})\equiv\delta\frac{\partial \mathbf{x}_{r}}{\partial\epsilon}\left(\mathbf{X}\right)+ h_{i}^{2} \mathbf{w}_{2i}\left(\mathbf{X}\right)+\sum_{j\ne i}h_{j}^{2}\tilde{\mathbf{u}}^{(2,0)}_{j}\left(\mathbf{X}\right)\label{eq:deformation-increment},
\end{equation} 
which is the deformation in the outer region of the slab, excluding the singular dipole term  due to the $i^{th}$ sulcus, $a\mathbf{f}_1(\theta)r_i^{-1}$. The terms on the right hand side of  Eq.  (\ref{eq:deformation-increment}) are, in order, the residual incremental deformation near $\mathbf{X}_i$  due to the change in $\epsilon$, the linear response of the system due to  the  growth of the sulcus at $\mathbf{X}_i$, and  the influence of  sulci at other points. 

Formally $\mathcal{U}_i$ must satisfy two conditions for  the local strain to be maximal with the value $\epsilon_c$. First, the residual deformation gradient must be a pure incremental rotation, i.e.
\begin{equation}
\mathbf{\Sigma}: \frac{\partial\mathcal{U}_i}{\partial \mathbf{X}}(\mathbf{X}_i)=\mathbf{0}.
\label{eq:bifurcation-equations}
\end{equation}
where $\mathbf{\Sigma}$ is  the rank-4 tensor of tangent moduli in the reference configuration \footnote{For energy (\ref{eq:neo-hookean-energy}), $\Sigma_{ijkl}B_{kl}=\delta_{ij}\mathrm{tr}(\mathbf{B})+p\,\mathrm{cof}_{ij}(\mathbf{B})$, 
for
 any matrix
 $\mathbf{B}$;
  at a free surface $p=(1-\epsilon)^{-2}$.}. Two components of 
 Eq. (\ref{eq:bifurcation-equations}) automatically vanish because $\mathbf{X}_i$ is on a free boundary. One additional component vanishes because of the objectivity of the constitutive equation. For a given choice of crease location points $\left\{\mathbf{X}_i\right\}_{i=1}^N$,  the substitution of Eq. (\ref{eq:deformation-increment}) into Eq. (\ref{eq:bifurcation-equations}) yields a linear system relating $h_i^2$ to $\delta$, which is solvable.  The points $\mathbf{X}_i$ are fixed by the second condition,
 \begin{equation}
\mathbf{\Sigma}:\frac{\partial^2 \mathcal{U}_i}{\partial X_1\partial\mathbf{X}}(\mathbf{X}_i)=\mathbf{0}\label{eq:translation-equation},
\end{equation}
which states that the local surface-parallel strain gradient vanishes. When Eqs. (\ref{eq:bifurcation-equations}) and (\ref{eq:translation-equation}) are satisfied, we find that $\mathbf{u}_i^{(2)}$ is a pure displacement and $\mathbf{u}_i^{(3)}$ is a pure rotation of the $i^{th}$ sulcus.
We can solve Eqs. (\ref{eq:bifurcation-equations}) and (\ref{eq:translation-equation})  to determine $\alpha$ in Eq.  (\ref{eq:h-eps-relation}) and find good agreement with our numerical simulations. For all pattern wavelengths $L$ in supported slabs, $\alpha>0$ and the bifurcations are supercritical [see Fig. (\ref{fig:fig1}c)]. For the unsupported slab,  sulci in each pattern  with  $L=2.4$ have identical depth, i.e. $h_i=h$. We find that Eq. (\ref{eq:h-eps-relation}) holds for these as well, but $\alpha<0$ and the bifurcations are subcritical [see Fig. (\ref{fig:fig2})]. 

In the case of a uniformly compressed slab, Eqs. (\ref{eq:bifurcation-equations}-\ref{eq:translation-equation}) only specify the pattern up to the number $N$ of sulci. However, as we discussed in the context of  simulations of supported slabs with periodic patterns, pattern selection at onset is controlled by  minimizing the reduced energy $\Delta \tilde{E}$ over trial patterns. The energy of a pattern can be divided into the work of forming and growing the sulcus cores and the energy of deforming the outer region. We find the energetic cost of forming a sulcus core is exactly canceled by the work done by the far field prestress, formally this implies
\begin{equation}
 \int_{\mathbb{R}^2_<}W\left(\mathbf{A}_{c}+\frac{\partial\mathbf{v}_s}{\partial \mathbf{X}}\right)-W(\mathbf{A}_{c})- \frac{\partial W}{\partial \mathbf{A}}(\mathbf{A}_{c}):\frac{\partial \mathbf{v}_s}{\partial \mathbf{X}} d^2\mathbf{X}=0,\label{eq:inner_energy}
\end{equation}
 where $\mathbb{R}^2_<\equiv \{X_2<0\}.$ The last term in the integrand of Eq. (\ref{eq:inner_energy}) renders the integral convergent because $|\mathbf{v}_s|\sim|\mathbf{Y}_i|^{-1}$.  In a finite domain, the formation of a sulcus core cancels the diverging self-energy integral for the corresponding force dipole in the outer region. Then  the only nonvanishing contribution to $\Delta \tilde{E}$ comes from the  remaining, convergent part of the energy integral (\ref{eq:neo-hookean-energy}) in the outer region. Values of  $\Delta \tilde{E}$ computed with our asymptotic theory agree with our simulations of supported slabs [see Fig (\ref{fig:fig1}a) (inset)] and unsupported slabs [see  Fig. (\ref{fig:fig2}) (inset)].

Our numerical simulations and asymptotic theory thus show that the scale and nature of sulcus patterns are determined by the linear response of the outer region to a short-wavelength instability, the formation of infinitesimal sulci at the points $\mathbf{X}_i$. Each infinitesimal sulcus breaks the scale and translation symmetries of a related auxiliary problem describing the formation of an isolated sulcus in a half-space. These broken symmetries manifest as secular growth in the series expansion of the solution and can result in new, soft bending modes of a critically strained slab such as the bucklinglike sulcus pattern shown in Fig. (\ref{fig:fig2}). The resulting bifurcation is nonlinear or supercritical depending on whether the response of the outer region increases or deceases the local compressive strain driving sulcification; in either case the sulci move and grow to remain at maxima of the local strain where the compression takes the critical value $\epsilon_c$. The physical reason for this behavior is that a sulcus has a finite energy of transformation,  formally stated by Eq. (\ref{eq:inner_energy}). However, sulcification is unlike familiar solid-solid phase transitions since the two ``phases'' here, the inner and outer regions, are not divided by a phase boundary. Although we have focused  on externally strained soft elastomers, our theory may also help illuminate  biological patterns of sulci and is likely to carry over more generally to other processes in elastomers with characteristic critical strains, e.g. cavitation and other Biot-like instabilities at interfaces \cite{Biot}.

\begin{acknowledgments} 
{We thank} Lev Truskinovsky for discussions.
\end{acknowledgments}

\bibliographystyle{plain}

\end{document}